\documentstyle[12pt,epsf]{article}
\newcommand{\lsim}{ \raisebox{-.5ex}{\mbox{$\,\stackrel{<}{\sim}$\,}} }
\begin{document}
\begin{center}
{\large\bf Dynamics of a Simple Model for Turbulence\\[2mm]
           of the Second Sound in Helium II}\\[2mm]

B.V. Chirikov\footnote{Email: chirikov @ inp.nsk.su} and V.G. Davidovsky\\[2mm]
{\it Budker Institute of Nuclear Physics \\
        630090 Novosibirsk, Russia}\\[5mm]
\end{center} 
\baselineskip=15pt

\vspace{2mm}

\begin{abstract} The results of numerical experiments on chaotic ('turbulent')
dynamics of the second sound in helium II are presented and discussed
based on a very simple model proposed and theoretically studied recently by
Khalatnikov and Kroyter. Using a powerful present--day
techniques for the studying nonlinear phenomena, we confirm their results on 
the stationary oscillation in helium and its stability as well as on
a qualitative picture of successive transitions to limit cycles and chaos.
However, the experiments revealed also a much more complicated
structure of the bifurcations than it was expected.
The fractal structure of chaotic
attractors was also studied including their noninteger dimension.
Surprisingly, a very simple model used in all these studies not only
qualitatively represents the behavior of helium in laboratory experiments
but also allows for a correct order--of--magnitude estimate of the critical
heat pumping into helium at bifurcations.
\end{abstract}

\section{Introduction}
The present studies were stimulated by the theoretical analysis of the
dynamics of the second sound in helium driven by the external heat periodic
perturbation \cite{1}.
In turn, this theory was developed to explain the experiments on the second
sound propagation in a liquid helium with a free surface \cite{2,3}.
Originally, such experiments were intended for a new scheme of holography
using the deformation of the helium surface under the standing waves
in the bulk of the liquid \cite{4}. To the best of our knowledge,
such a project was never realized. The reason is very simple: a negligible
surface deformation which is determined by the oscillation of the pressure 
in a second--sound
wave, which is much smaller than the temperature oscillation.
Nevertheless, the experiments were continued, and an interesting dynamics
was observed, namely the transition, at a sufficiently high heat pumping
into the helium, from a stationary second--sound wave to its chaotic motion,
or turbulence \cite{3}. This was just the subject of the theoretical analysis
in Ref.\cite{1}. The latter was necessarily restricted, even in the simplest 
model considered, to the calculation of the stability of a stationary
standing wave. For this reason the authors \cite{1} began also some numerical
experiments which qualitatively confirmed their conjecture on a transition
to limit cycles and chaos. 
The main purpose of our numerical studies is to continue such experiments
for a more detailed and reliable analysis of those transitions.

\section{The model}
We adopted the model developed in Ref.\cite{1} which can be specified by the
following effective non--Hermitian Hamiltonian (in notations of Ref.\cite{1}
with only a few minor modifications):
$$
   H(a_1,\,a_2,\,t)\,=\,(\omega_1\,-\,i\gamma_1)|a_1|^2\,+\,
   (\omega_2\,-\,i\gamma_2)|a_2|^2\,
   +\,\left(\lambda a_1^2a_2^*\,+\,fa_1^*{\rm e}^{-i\omega t}\,
   +\,{\rm c.c.}\right) \eqno (2.1)
$$
This Hamiltonian describes the two wave modes (oscillators) via complex 
phase--space variables $a_j\ (j=1,2)$ and the complex small--oscillation 
frequencies $\omega_j-i\gamma_j$
with phenomenological dissipation parameters $\gamma_j$.
Two other parameters of the model, representing some effective
nonlinear coupling of the two modes $\lambda$ and the driving
'force' (external heat pumping) $f$ of frequency $\omega$,
can be chosen real and positive.
The model parameter $f$ is related to the input heat power by an approximate 
expression
$$
   P\,\approx\,\frac{\partial H}{\partial t}\,=\,
   -2f\omega\,{\rm Im}(\bar{a}_1)
   \eqno (2.2)   
$$
provided the parameters $\gamma_j$ and $f$ are sufficiently small.
Here $\bar{a}_j$ are the new variables ('slow' amplitudes of the wave modes)
determined by the canonical transformation
$$
\begin{array}{lll}
 a_1 & = &\bar{a}_1 e^{-i \omega t} \\
 a_2 & = &\bar{a}_2 e^{-2i \omega t}
\end{array}
\eqno (2.3)
$$
which eliminates the explicit dependence on time in Eq.(2.1).
Below we omit the upper bar in all $a_j$.
Notice that the new Hamiltonian still exactly describes the original
model (2.1).
The corresponding motion equations have the form:
$$
\begin{array}{lll}
 i \dot a_1 & = & (\Delta_1-i \gamma_1) a_1+2\lambda a_1^* a_2+f \\
 i \dot a_2 & = & (\Delta_2-i \gamma_2) a_2+\lambda a_1^2
\end{array}
\eqno (2.4)
$$
where $\Delta_1=\omega_1-\omega$ and $\Delta_2=\omega_2-2\omega$
are the detunings of the small--amplitude oscillation frequencies with respect 
to that of the driving force.

The described model is actually the simplest possible one for representing
a complicated dynamics of the real physical system under consideration.
Particularly, only the lowest (third--order in Hamiltonian) nonlinear term 
is included.
Meanwhile, the next, fourth--order term is known to give a comparable
contribution, particularly to the nonlinear shift of the oscillation
frequencies. In Ref.\cite{1} an argument is given for neglecting this
term. In our opinion, the main argument is in that the next term does not
lead to new qualitative effects, and moreover retains the order--of--magnitude
of the description which is the best one could expect from such a
primitive model.
In any event, following
Ref.\cite{1} we also restrict our studies to the equations (2.4).

\section{Numerics}
In numerical experiments the set of equations (2.4) was integrated
by itself (for trajectories) or together with the linearized
equations (for Lyapunov exponents) using the Fehlberg version of the 
forth--order
Runge - Kutta algorithm. A typical integration step was
$\Delta t\sim 2^{-18}$. The accuracy of integration was checked,
as usual, by a change of the step size.
In case of a stable motion (stationary oscillation, limit cycle) 
the criterion of accuracy was a negligible change of the trajectory.
However, for chaotic motion this was never possible, and we checked
instead the accuracy of statistical properties of the motion, for example of
the Lyapunov exponents.

We computed all (4) exponents 
$\Lambda_n\ (\Lambda_{max}\equiv\Lambda_1\geq\Lambda_2\geq\Lambda_3
\geq\Lambda_4$) 
using also
the standard method (see, e.g., Ref.\cite{5}).
One of them, whose eigenvector goes along the 
trajectory, is always zero while the sum of all 
$$
   \Lambda_{\Gamma}\,=\,\sum_{n=1}^4\,\Lambda_n\,=\,
   -2(\gamma_1\,+\,\gamma_2)\,=\,
   {\rm const} \eqno (3.1)
$$ 
is the rate of the phase space volume contraction.
Both conditions were also used to check the integration accuracy.

All the model parameters but $f$ were fixed to (in CGS units)
$$
   \begin{array}{ll}
   \Delta_1=0 &\Delta_2=-1500 \\
   \gamma_1=30 &\gamma_2=120 \\
   \lambda =5400 &{\rm and}\\
   \omega = 8000
   \end{array}
   \eqno (3.2)
$$
as in Ref.\cite{1}. The principal parameter $\lambda$ was calculated
in Ref.\cite{6}. The rest were chosen close to those in a typical laboratory
experiment \cite{2,3}. We add also the driving frequency $\omega$ \cite{2},
absent in Eq.(2.4), which we will need below for a quantitative comparison
with the laboratory experiments.
The remaining parameter $f$, related to the input heat power (see Eq.(2.2)), 
was the only varying one in this first series of our
numerical experiments.

In agreement with laboratory experiments the initial conditions were fixed at
$$
   a_1(0)\,=\,a_2(0)\,=\,0 \eqno (3.3)
$$
which is important for the interpretation of numerical experiments below.

The main results of our experiments are presented in Fig.1
as a dependence of the maximal Lyapunov exponent 
$\Lambda_{max}(f)$
on the driving parameter $f$.

Three qualitatively different regimes of the motion are 
clearly seen:
\begin{enumerate}
\item[(i)] $\Lambda_{max}<0$ and, hence, all $\Lambda$'s are
         negative. This means a stationary standing wave
         according to the original Hamiltonian (2.1) or
         the constant amplitudes ($a_j=$const)  in the reduced
         motion equations (2.4) that is a fixed point in the
         amplitude space. This is the simplest regime well
         recognized in both laboratory and numerical experiments.
\item[(ii)] $\Lambda_{max}=0$.
          This may be a periodic (limit cycle) or quasiperiodic attractor.
          In the former (simplest) case the trajectory is closed.
          One way to recognize
          the cycle is simply to see a picture of the trajectory, or
          rather of some two--dimensional representation
          of the four--dimensional trajectory. This was done
          in Ref.\cite{1}, and we will not repeat it here.
          However, in case of a quasiperiodic attractor this visual
          analysis doesn't work.
\item[(iii)] $\Lambda_{max}>0$. This is the most simple and reliable criterion
             for chaotic motion. The visual difference between the chaotic
             attractor and a complicated limit cycle is not always clear
             and may be deceptive.
\end{enumerate}

\begin{figure}[h]
\centerline{\epsfxsize=15cm \epsfbox{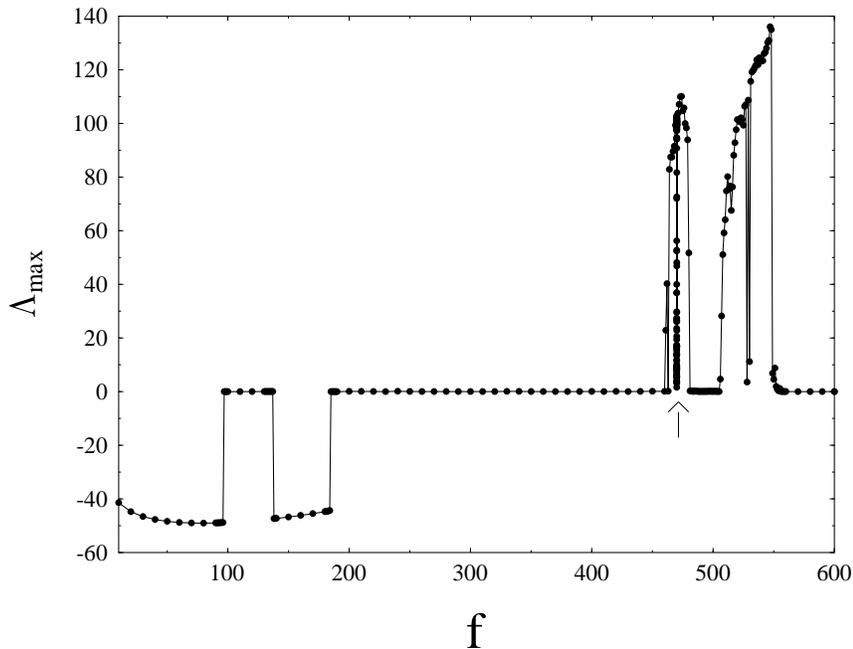}}
\caption{The global dependence of the maximal Lyapunov exponent $\Lambda_{max}$
         on the driving force $f$ revealing the three types of attractor
         in model (2.4) with parameters (3.2) and initial conditions (3.3).
         Empirical points are connected by line to guide the eye.}
\end{figure}

\begin{figure}[h]
\centerline{\epsfxsize=15cm \epsfbox{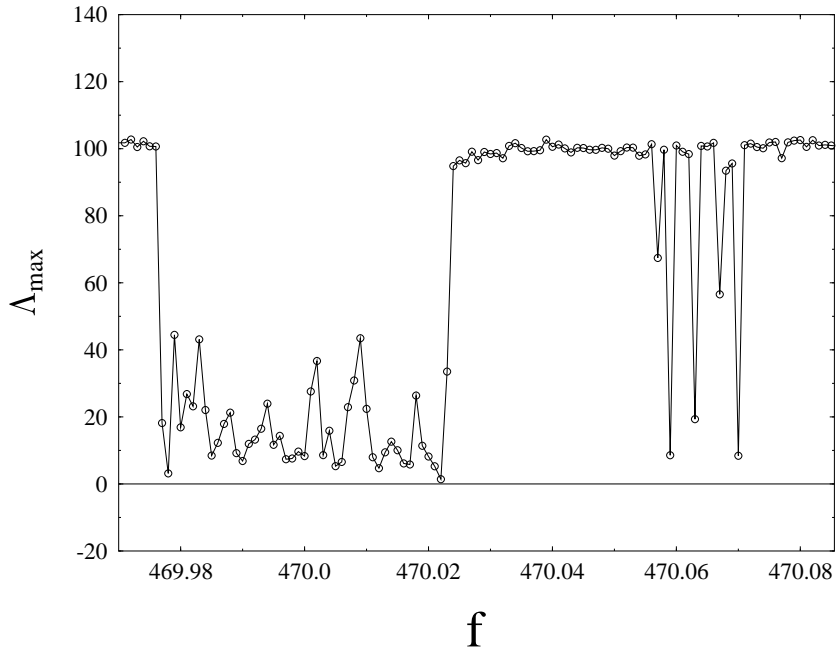}}
\caption{Same as in Fig.1 for  a very narrow interval of $f$,
          indicated by arrow, where a deep drop of $\Lambda_{max}>0$
          occurs.
}
\end{figure}
           
As the force $f$ grows the first very sharp bifurcation 
(fixed point/limit cycle)
occurs at $f_{1}\approx 97$ but it does not mean the fixed--point attractor
disappears for larger $f$. Indeed, both the theory \cite{1} and our
numerical experiments show that the attractor retains up to $f\approx 380$
for some initial conditions of the motion which are generally different
from (3.3). What actually happens in the first bifurcation is the escape
of the initial point (3.3) out of the basin of attraction \cite{1}.
Particularly, this explains the 'window' with a fixed--point attractor 
at larger $f\approx 137\ -\ 185$ (Fig.1).

At still greater $f=f_{2}\approx 461$ a new and much more interesting 
bifurcation occurs (limit cycle/chaotic attractor). Actually, it is a
complicated chain of many bifurcations between periodic and chaotic
attractors. This is especially clear in Fig.2 where a narrow part of
dependence $\Lambda_{max}(f)$ is shown in detail (cf. Fig.1).
 
The structure of each chaotic attractor is rather intricate, and particularly
fractal. An important characteristic of such a structure is the dimension
which is generally noninteger (see, e.g., Ref.\cite{5}).
The dimension can be calculated from the Lyapunov exponents
using the relation
$$
   d\,=\,m\,+\,\frac{\sum_{n=1}^m\,\Lambda_n}{|\Lambda_{m+1}|}
   \eqno (3.4)
$$
where $m$ is the largest integer for which the sum $S_m=\sum_{n=1}^m \Lambda_n
\geq 0$. This is the simplest and widely used method for calculating
the dimension.
For example, on a fixed--point attractor (all $\Lambda$'s are negative)
$m=0$, and $d=0$. Similarly, for a periodic attractor (limit cycle)
$m=1$, and $d=1$ (a closed trajectory). In case of a quasiperiodic attractor
$m=d=2\ {\rm or}\ 3$, so that the trajectory is asymptotically restricted
to a phase--space surface of dimension 2 or 3, respectively.
We tried but couldn't find any quasiperiodic attractor in the model under
cosideration.

Unlike this, the dimension of a chaotic attractor is fractal because two
of the four Lyapunov exponents are independent. In principle,
three different cases are 
possible: (i) $\Lambda_2=0\  (\Lambda_1>0)$;
(ii) $\Lambda_3=0\ (\Lambda_1\geq\Lambda_2>0)$;
and (iii) $\Lambda_2=
\Lambda_3=0\ (\Lambda_1>0)$. In our experiments we have 
always seen the first case only and never the two latter ones.

In the experiments presented in Fig.1 for $f\leq 600\ (\Lambda_1\leq 140$
the attractor dimension was in the interval
$(2<d<3)$ corresponding to $m=2$ and $S_m=\Lambda_1$ (see Eq.(3.4)).
For larger $f$ the dimension grows due to increase of $m$
($S_m=S_3=\Lambda_1+\Lambda_3\geq 0$), 
and can be represented in the form:
$$
   d(S_3)\,=\,3\,-\,\frac{S_3}{\Lambda_4}\,=\,3\,+\,
   \frac{S_3}{|\Lambda_{\Gamma}|\,+\,S_3}\,<\,4 \eqno (3.5)
$$
In agreement with the physical meaning of the dimension it never exceeds
the upper bound $d_{max}=4$. Whether $d$ is actually approaching this bound is
an interesting question. However, it makes sense for the model only,
since at such a big $f$ the model becomes
certainly inadequate for the real physical system under consideration.

The above analysis of attractors in the model was confirmed by the computation
of the motion spectra and correlation functions.

\section{Discussion}
How strange it may seem the very simple Khalatnikov - Kroyter
model considered above does describe
some qualitative results of the laboratory experiments \cite{2,3}.
The most important of those is the observed transition from
a stationary standing wave to something which looks like a
chaotic motion or wave turbulence.

In numerical experiments three types of attractors are clearly
seen (Fig.1): the stationary standing wave, limit cycles, 
and chaotic attractors.
In laboratory experiments it is generally rather difficult 
to be sure that a particular attractor is truly chaotic one
rather than a very complicated limit cycle, for example.
Nevertheless, the experimental observations presented in Ref.\cite{2,3}
provide some plausible indications of the transition to turbulence.
The most clear evidence of that can be found in Fig.2 \cite{3}
where the motion of the helium free surface is shown for a few (3) values
of the input heat power. According to these data, the first transition
(stable surface/limit cycle) occurs at a power $P_{1}\lsim 18$ mW.
The inequality is explained by the absence of experimental data for
lower power, so it remains unclear whether the transition at
$P_{1}\approx 18$ mW is the first escape of the system out of the
basin of attraction or it is a window at a higher power like in our Fig.1
for the model.
The second (most interesting) transition to an apparent turbulence
takes place somewhere in the interval $P_{2}\approx 18\ -\ 93$ mW.
Here we take the lower $P_2$ equal to the upper $P_1$ for the same reason,
the absence of intermediate experimental data.

In the model, the critical power input can be estimated from Eq.(2.2) as
$$
   P_{1}\,\approx\,2.6\ {\rm mW} \quad {\rm at} \quad f\,\approx\,95,
   \ {\rm Im}(a_1)\,\approx\,-0.016 \eqno (4.1a)
$$
for the first transition and the first escape out of the attraction basin 
(see Fig.1) or
$$
   P_{1}\,\approx\,4.8\ {\rm mW} \quad {\rm at} \quad f\,\approx\,170,
   \ {\rm Im}(a_1)\,\approx\,-0.018 \eqno (4.1b)
$$
for the second escape.
As to the transition to a chaotic attractor, the critical power 
in the model is 
$$
   P_{2}\,\approx\,26\ {\rm mW} \quad {\rm at} \quad f\,\approx\,450,
   \ {\rm Im}(a_1)\,\approx\,-0.036 \eqno (4.1c)
$$
In such a simple model
it seems to be a quite reasonable agreement 
(especially for the transition to turbulence), 
and in view of a considerable uncertainty
in laboratory experiments.

The success of the Khalatnikov - Kroyter model can be explained, at least
partly, by the special choice of the parameters of experiments when
one of the wave modes was in resonance with the input heat oscillation.
Since the dissipation in higher modes is rapidly increasing, only a few
lower modes are excited initially which turned out to be already sufficient 
for the transition to turbulence. For a higher power input
the number of excited modes also rapidly grows, and the model becomes
inadequate.

An interesting question about the maximal dimension of a chaotic attractor
in the model (see Section 3), being irrelevant for the 
physical model under consideration here, may still be of importance
for the theory of dissipative dynamical systems, and deserves
further studies.

\vspace{5mm}

{\bf Acknowledgments.} We are grateful to I.M. Khalatnikov for attracting
our attention to this interesting physical problem, and for providing 
an opportunity
to acquaint ourselves with his and M. Kroyter's results prior to publication.
We appreciate many interesting and stimulating discussions with both of them.


\end{document}